\documentclass[11pt]{article}

\def \noi{\noindent}

\def \ssk{\smallskip} 

\def \aa {{\it Astron. Astrophys.}}

\def \jgr {{\it J. Geophys. Res.}}

\usepackage{latexsym}
\usepackage{amsmath}
\usepackage{amssymb}
\usepackage{amsfonts}
\usepackage{bm}

\usepackage{graphicx}

\begin{document}

\vspace*{1.2cm} 

\noi {\Large GRAVITATIONAL POTENTIAL, INERTIA AND EARTH ROTATION} 

\vspace*{1cm} 

\noi \hspace*{1.5cm} G\'{e}raldine BOURDA 

\noi \hspace*{1.5cm} SYRTE - UMR8630/CNRS, Observatoire de Paris  

\noi \hspace*{1.5cm} 61 avenue de l'Observatoire - 75014 Paris, FRANCE 

\noi \hspace*{1.5cm} e-mail: Geraldine.Bourda@obspm.fr 

\vspace*{1cm}


\noi {\large INTRODUCTION} 

\ssk
\noi Several satellite missions, devoted to the study
of the Earth gravity field, have been launched (like CHAMP, recently). This year,
GRACE (Gravity Recovery and Climate Experiment) will allow us to obtain 
a more precise geoid. But the most important is that they will supply the temporal 
variations of the geopotential coefficients (called Stokes coefficients). 

\noi In the poster, we show how the Earth gravitational potential is linked 
to the Earth rotation parameters.
Indeed, through the Earth inertia coefficients, we can connect the variation of LOD and Polar Motion with 
the temporal variations of the Stokes coefficients.
We also consider the nutations, that are related to the gravitational geopotential coefficients.

\noi We discuss the possibility of using the Stokes coefficients in order to improve 
our knowledge of the Earth rotation.

\vspace*{1cm} 


\noi {\large 1. LENGTH-OF-DAY} 

\ssk
\noi The excess in the length-of-day can be related to
the instantaneous Earth rotation rate $\omega = \Omega~(1+m_3)$ 
, where $\Omega$ is the mean Earth speed of rotation.
\noi Indeed, we have :
\begin{equation}
\Delta(LOD) = LOD - LOD_{mean} = k~\frac{2 \pi}{\omega} -  k~\frac{2 \pi}{\Omega} 
\simeq -k~\frac{2 \pi}{\Omega} ~m_3
\end{equation}
where $k$ is the conversion factor from sidereal to mean solar days. That involves :
\begin{equation}
- \frac{\Delta(LOD)}{LOD_{mean}} = m_3 
\end{equation}
By the way of the Liouville's equations, $m_3$ is linked with $L_3$,
third component of the external torque, with $c_{33}$, time-dependant difference to the constant
part $C$ of the third principal moment of Earth inertia and with $h_3$, third component
of the relative angular momentum of the system. In this case, not considering 
the external perturbations, we have :
\begin{equation}\label{eq:LOD_c33_h3}
\frac{\Delta(LOD)}{LOD_{mean}} =  \frac{c_{33}}{C}  +  \frac{h_3}{C~\Omega}  
\end{equation}
Furthermore, the variable part $c_{33}$ of the Earth inertia tensor 
depends on the temporal variation of the Stockes coefficient $C_{20}$ of degree 2 and order 0 
(Gross, 2000) :
\begin{equation}\label{eq:c33}
c_{33} (t) = \frac{1}{3}~\Delta Tr(I) - \frac{2}{3} ~\mathcal{M}~\mathcal{R}_e^2 ~\Delta C_{20} (t)
\end{equation}
where $\Delta Tr(I)$ is the time-dependent difference to the constant part of the inertia tensor trace.
Due to the conditions in which we place (fluid atmospheric and oceanic layers parts of our system),
we can consider that there is conservation of the volume of our system under deformations.
For this reason, according to (Rochester and Smylie, 1974), we have : $\Delta Tr(I)=0$.
Finally, according to the equations (\ref{eq:LOD_c33_h3}) and (\ref{eq:c33}), 
but also considering surface loading and rotationnal deformation (Barnes et al., 1983)
(factor 0.7), we obtain :
\begin{equation}\label{eq:LOD_fin}
\frac{\Delta(LOD)}{LOD_{moyen}} =  
- 0.7 ~\frac{2}{3 ~C} ~\mathcal{M}~\mathcal{R}_e^2 ~\Delta C_{20} +  \frac{h_3}{C~\Omega}  
\end{equation}

\vspace*{1cm} 


\noi {\large 2. POLAR MOTION} 

\ssk
\noi According to (Gross, 1992), we can connect the polar motion $p=x-iy$
with the motion of the instantaneous rotation axis $m = m_1 + i ~m_2$ :
$ m = p - i/\Omega ~\dot{p}$.
Considering (Gross, 2000), this brings to link the polar motion to the temporal 
variations of some Stockes coefficients :
$$
\left\{
\begin{array}{l}
p + i~\frac{\dot{p}}{\sigma_r} = \chi  \\
\chi  = \frac{1}{\Omega~(C-A)} (\Omega~c+h) 
= \frac{1}{\Omega~(C-A)} \left( -\mathcal{M}~\mathcal{R}_e^2~\Omega ~(C_{21}+i~S_{21}) + 1.43 ~h \right)  
\end{array}
\right.
$$
where $C_{21} \equiv \Delta C_{21}$, $S_{21} \equiv \Delta S_{21}$, $c=c_{13} + i ~c_{23}$,
and the coefficient $1.43$ comes from (Barnes et al., 1983).

\vspace*{0.8cm} 


\noi {\large 3. NUTATIONS} 

\ssk
\noi Some authors have derived the nutations of the Earth rotation axis
using the harmonic coefficients of the gravity geopotential (Melchior,1973; Bretagnon, 1997; 
Brezinski and Capitaine, 2001).

\noi The most important Stokes coefficient for such a phonomenon is $C_{20}$, but
we can see that the other zonal coefficients act on the nutations. It would be then interesting 
to have their temporal variations, even if they are diurnal in the Earth.

\vspace*{0.8cm} 


\noi {\large DISCUSSION} 

\ssk
\noi We have provided the equations linking the temporal variations of the geopotential coefficients 
with the Earth rotation parameters (variation of LOD, polar motion and nutations). 
With the new gravity missions, we will be able to determine variations of Stokes coefficients,
and to isolate the total variations of the solid Earth moment of inertia, what is totally new.


\noi This is a new method, so comparing our results with those already known will
be interesting, but a study of the precision needed for such a work is necesary 
and will be done.

\vspace*{0.8cm} 


\noi {\large REFERENCES}

\ssk
\noi Barnes R., Hide R., White A., Wilson C., 1983, \textit{Proc. R. Soc.} pp 31-73

\noi Bretagnon P., Rocher P., Simon J., 1997, \aa  pp 305-317

\noi Brzezi\'{n}ski A., Capitaine N., 2001, \textit{Proceedings Journ\'{e}es 2001 Syst\`{e}mes de R\'{e}f\'{e}rences
spatio-temporels} pp 51-58, N. Capitaine (ed.), Observatoire de Paris 

\noi Gross R., 1992, \textit{Geophys. J. Int.} pp 162-170

\noi Gross R., 2000, \textit{Gravity, Geoid and Geodynamics 2000}, IAG Symposium 123


\noi Lambeck K., 1988, "Geophysical geodesy : The slow deformations of the Earth", Oxford Science Publications

\noi Melchior P., 1973, "Physique et dynamique plan\'{e}taires : G\'{e}odynamique", Vol. 4, Vander

\noi Rochester M., Smylie D., 1974, \jgr pp 4948-4951


\end{document}